\documentclass[fleqn,10pt]{wlscirep}

\usepackage{ulem}
\usepackage{color}

\title{Efficient Scheme for Perfect Collective Einstein-Podolsky-Rosen Steering}
\author[1,2]{M. Wang}
\author[1,2]{Q. H. Gong}
\author[3]{Z. Ficek}
\author[1,2,*]{Q. Y. He}
\affil[1]{State Key Laboratory of Mesoscopic Physics, School of Physics, Peking University, Beijing 100871, P. R. China}
\affil[2]{Collaborative Innovation Center of Quantum Matter, Beijing 100871, P. R. China}
\affil[3]{The National Center for Applied Physics, KACST, P.O. Box 6086, Riyadh 11442, Saudi Arabia}
\affil[*]{qiongyihe@pku.edu.cn}

\begin{abstract}
A practical scheme for the demonstration of perfect one-sided device-independent quantum secret sharing is proposed. The scheme involves a three-mode optomechanical system in which a pair of independent cavity modes is driven by short laser pulses and interact with a movable mirror. We demonstrate that by tuning the laser frequency to the blue (anti-Stokes) sideband of the average frequency of the cavity modes, the modes become mutually coherent and then may collectively steer the mirror mode to a perfect Einstein-Podolsky-Rosen state. The scheme is shown to be experimentally feasible, it is robust against the frequency difference between the modes, mechanical thermal noise and damping, and coupling strengths of the cavity modes to the mirror.
\end{abstract}
\begin{document}

\flushbottom
\maketitle

\thispagestyle{empty}

\section*{Introduction}
Quantum steering~\cite{wj07,cj09} is currently attracting considerable theoretical and experimental interest. The term steering was introduced by Schr{\"o}dinger~\cite{es35} for the fact that entanglement would  allow an experimentalist to remotely steer or pilot the state of a distant system as considered in the original Einstein-Podolsky-Rosen (EPR) paradox~\cite{ep35}. The distinctive feature of quantum steering is its directionality in a sense that if two parties, Alice and Bob, share an entangled state then the measurements made on the Bob's system can remotely affect, i.e. steer the Alice's system to a specific state. This makes quantum steering an essential resource for a number of applications, such as quantum key distribution~\cite{bc12,ww14}, secure quantum teleportation~\cite{md13} and performing entanglement assisted subchannel discrimination~\cite{pw14}. Quantum steering allows two parties to verify the shared entanglement even if one measurement device is untrusted~\cite{Opanchuk}. The idea of quantum steering has been theoretically
investigated~\cite{spw11,sd13,olsen13,chen13,hf14,bv14,sn14,gsa14,tr14,wg14,wgh14,kim14,he15,ade14,ade15} and also experimentally tested for several
systems~\cite{sj10,sg12,be12,vh12,bw12,ss13,Guo14,nc15,Pan15,seiji15}.

Recently, the concept of {\it collective} steering has been introduced in multipartite systems~\cite{hr13},  where the steering of a system $i$ by a group of $N-1$ parties cannot be demonstrated by the measurement of $N-2$ or fewer parties. This means the measurements made on $N-1$ parties enable them collectively steer the quantum state of a given party $i$, i.e. can infer the position and the momentum of the mode $i$ with higher precision than that allowed by the level of quantum standard limit, whereas a measurement on $N-2$ or fewer modes cannot infer this information. This feature thus opens the possibility for the realization of more secure multi-mode quantum cryptography, such as quantum secret sharing (QSS)~\cite{QSS}. QSS aims at protecting a highly important message, by demanding that all receivers must collaborate to decrypt the secret sent by the sender. Unlike the conventional QSS protocols, the usage of the collective steering as quantum resource need not assume the collaborating parties are trustworthy~\cite{bc12}, where the security of this process relies on the intrinsic nature of collective steering. Thus, it significantly reduces the device requirements in the network, and may provide unique conceptual tools for one-sided, device-independent quantum secret sharing (1SDI-QSS)~\cite{wgh14,seiji15}.

\begin{figure}[ht]
\centering
\includegraphics[width=12cm]{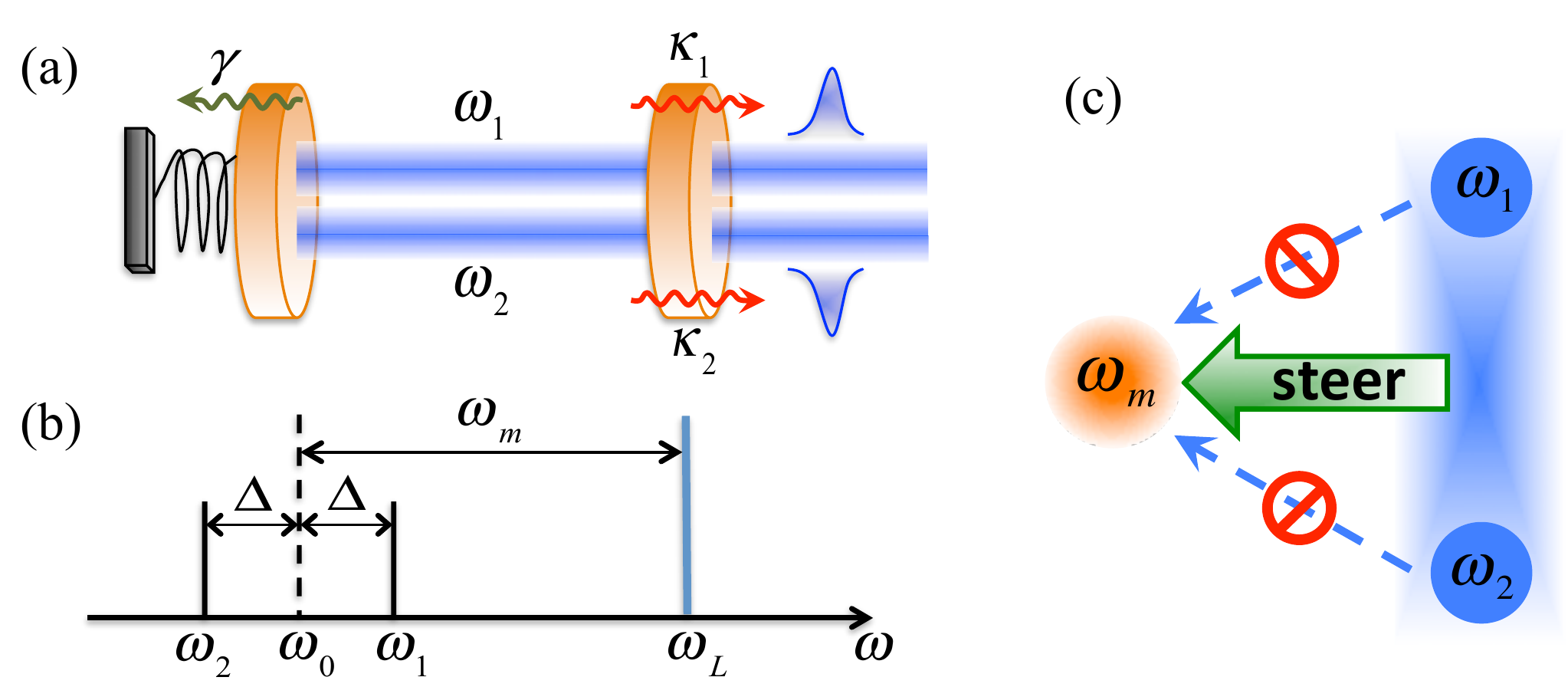}
\caption{\textbf{Schematic diagram of a three-mode optomechanical system for realization of perfect EPR collective steering.} (a) The system consists of two nondegenerate cavity modes of frequencies $\omega_{1}$ and $\omega_{2}$, separated by $2\Delta=\omega_{1}-\omega_{2}$, and a single mode of frequency $\omega_{m}$ associated with the oscillating mirror. The cavity modes are driven by a pulse laser with duration time $\tau$ and are damped with rates $\kappa_{1}$ and $\kappa_{2}$, respectively. The damping rate of the vibrating mirror is $\gamma$. (b) The laser frequency $\omega_{L}= \omega_{0}+\omega_{m}$ is tuned to the blue (anti-Stokes) sideband of the average frequency $\omega_{0}$ of the cavity modes. (c) Two cavity modes can collectively steer the quantum state of the oscillator (indicated by green arrow), i.e. can infer the amplitude of the mirror to below the level of quantum standard limit in both position and momentum, while the cavity modes cannot steer the mirror individually (indicated by red stop sign).}
\label{fig1}
\end{figure}

In this report, we propose a practical scheme to create the perfect collective steering of a macroscopic object in the context of pulsed cavity optomechanics~\cite{pnas11,hofer11}.  The scheme does not involve any external sources of squeezed light and networks of beam splitters used in the linear optical schemes~\cite{wgh14,seiji15}. The scheme consists of a pair of nondegenerate cavity modes driven by detuned short laser pulses and interacting collectively with a bosonic mode associated with the movable mirror, as illustrated in Fig.~\ref{fig1}(a). We show that the action of measuring the output of the two cavity fields can steer the quantum state of the oscillator and thus enable error-free predictions for its position and momentum given some type of measurement on the cavity modes. The fundamental significance of observing the ``steering" of the quantum state of the oscillator can address the original EPR paradox for a macroscopic and massive object, which is more useful for insights about quantum effects with matter.

Notice that a pulsed scheme does not require a steady state to be applicable, and entanglement can be achieved without stability requirements~\cite{pnas11,hofer11}. The existing proposals for the generation of tripartite entangled states and quantum steering have generally been based on the two-mode coupling with the laser frequency tuned to the blue (anti-Stokes) sideband of one of the modes and to the red (Stokes) sideband of the other mode~\cite{milburn11,ke13,lin13,wang13,wang15}. The coupling results in the presence of both tripartite and bipartite steering of a given mode and as such it rules out the aspect of collective steering. We find that by putting the driving laser to the blue sideband of the average frequency of the cavity modes, (Fig.~\ref{fig1}(b)), one can create a tripartite steering without creating the bipartite steering of the mirror mode, as illustrated in Fig. \ref{fig1}(c). The only requirement for the absence of the bipartite steering is a uniform coupling of the cavity modes to the mirror mode. We then have a prototype of the collective steering experiment in which the macroscopic mirror and the collective mode of the two optical fields could be in a perfect EPR state. By analyzing realistic conditions including the mechanical thermal noise and damping, we estimate that collective steering can be readily achieved with existing experimental parameters~\cite{chan11,lehnert13}. This offers an efficient scheme for the realization of a perfect one-sided device-independent quantum secret sharing or more secure multi-mode quantum cryptography.

\section*{Results}
\textbf{Dynamics of the system.} We model the system by using the the standard Langevin formalism in which we include photon losses in the cavity and the Brownian noise acting on the mirror~\cite{gm09,pm12}.
We focus on the case where the laser frequency $\omega_{L}$ is on resonance with the blue sideband of the average frequency of the cavity modes, i.e., $\omega_{L}=\omega_{0}+\omega_{m}$, with $\omega_{0} = (\omega_1 + \omega_2)/2$. Then, we introduce fluctuation operators of the cavity modes and the mirror mode, and following the standard linearization method we find that under the rotating-wave approximation the operators satisfy coupled equations
\begin{align}
\delta\dot{a}_{1} &=  -\left(\kappa_1+i\Delta\right)\delta a_{1}-ig_{1}\delta {b}^{\dag}_{m}-\sqrt{2\kappa_1}a_{1}^{\rm in} ,\nonumber \\
\delta\dot{a}_{2} &=  -\left(\kappa_2-i\Delta\right)\delta a_{2}-ig_{2}\delta {b}^{\dag}_{m}-\sqrt{2\kappa_2}a_{2}^{\rm in} ,\nonumber \\
\delta\dot{b}_{m} &= -\gamma \delta{b}_{m} -ig_{1}\delta a_{1}^{\dag}-ig_{2}\delta a_{2}^{\dag}-\sqrt{2\gamma}b^{\rm in}_{m},\label{e2}
\end{align}
where $\delta{a}_{j}$ $(j=1,2)$  and  $\delta{b}_{m}$ are slowly varying parts of the fluctuation operators,  $\kappa_j$ is the damping rate of the cavity mode $j$, assumed the same for both modes,  $\kappa_1=\kappa_2=\kappa$, $g_{j}$ is the effective coupling strength of the $j$th cavity mode to the mirror, $\Delta =(\omega_{1}-\omega_{2})/2$, and $\gamma$ is the damping rate of the mirror. $a_{j}^{\rm in}$ and $b^{\rm in}_{m}$ are Langevin noise terms which are taken to be statistically independent with nonzero $\delta$ correlated functions, $\langle a_{j}^{\rm in}(t)a_{j}^{\rm in\dag}(t^{\prime})\rangle =\delta(t-t^{\prime})$ and $\langle b^{\rm in}_{m}(t)b^{\rm in \dag}_{m}(t^{\prime})\rangle = (n+1)\delta(t-t^{\prime})$, where~$n$ is the mean number of thermal phonons. To simplify the notation, we will drop the label $\delta$  on the fluctuation operators.

The solution of equation~(\ref{e2}) is in general complicated. A simple analytical solution arises, however, in two cases, the bad cavity limit, $\kappa \gg g_{1,2}$, or at a large difference between the cavity frequencies, $\Delta \gg g_{1,2}$, at which
\begin{align}
a_{j} = -\frac{e^{(-1)^{j}i\phi}}{\sqrt{\kappa^{2}+\Delta^{2}}}(ig_{j}{b}^{\dag}_{m} + \sqrt{2\kappa}a_{j}^{\rm in}),\label{e3}
\end{align}
where $\phi=\arctan\left(\Delta/\kappa\right)$. Then the equation of motion for the mirror mode becomes
\begin{align}
\dot{{b}}_{m} &= \left(G+i\delta\right){b}_{m} + i\sqrt{2G_{1}}e^{i\phi}a^{{\rm in} \dag}_{1}+i\sqrt{2G_{2}}e^{-i\phi}a^{{\rm in}\dag}_{2} -\sqrt{2\gamma}b^{\rm in}_{m} , \label{e4}
\end{align}
where $G_{j} ={g_{j}^{2}\kappa}/(\kappa^{2}\!+\!\Delta^{2})$, $\delta = (g_{1}^{2}- g_{2}^{2})\Delta/(\kappa^{2}\!+\!\Delta^{2})$, and $G = G_{1}+G_{2}-\gamma$.

\textbf{Collective cavity modes.} In order to calculate the fluctuation operators at an arbitrary pulse duration time $\tau$, we introduce normalized temporal pulse-shape amplitudes
\begin{align}
 A_{m}^{\rm out} &={b}_{m}(\tau)e^{-i\delta \tau} ,\quad A_{m}^{\rm in} ={b}_{m}(0) ,\nonumber \\
A_{j}^{\rm in} & =  e^{(-1)^{j}i\phi}\sqrt{\frac{2G}{1-e^{-2G\tau}}}\int_{0}^{\tau}dt\, a_{j}^{\rm in}(t)e^{-(G-i\delta)t} ,\nonumber \\
B_{m} &= \sqrt{\frac{2G}{1-e^{-2G\tau}}}\int_{0}^{\tau}dt\, b^{\rm in}_{m}(t)e^{-(G+i\delta)t} ,\label{e6}
\end{align}
and find that the output amplitude of the mirror mode $A_{m}^{\rm out}$ is only affected by the collective ``symmetric'' cavity mode
$W_{\rm in} = (\sqrt{G_{1}}A_{1}^{\rm in}\!+\!\sqrt{G_{2}}A_{2}^{\rm in}\!-i\sqrt{\gamma}B_{m}^{\dag})/\sqrt{G}$ via
\begin{eqnarray}
A_{m}^{\rm out} = A^{\rm in}_{m}e^{Gt} + i\sqrt{e^{2Gt}-1}\,W_{\rm in}^{\dag}.\label{e7}
\end{eqnarray}
This indicates that the cavity modes interact collectively rather than individually with the mirror mode. Note that the called ``output" mode for the mirror is the amplitude of the mirror mode at the final time $\tau$.

We then define the normalized pulse-shape amplitudes of the modes
\begin{align}
A_{j}^{\rm out} &= e^{-(-1)^{j}i\phi}\sqrt{\frac{2G}{e^{2G\tau}-1}}\int_{0}^{\tau}dt\, a_{j}^{\rm out}(t)e^{(G+i\delta)t} ,\nonumber \\
\tilde{A}_{j}^{\rm in} & = e^{(-1)^{j}i\phi}\sqrt{\frac{2G}{e^{2G\tau}-1}}\int_{0}^{\tau}dt\, a_{j}^{\rm in}(t)e^{(G+i\delta)t} ,\nonumber \\
\tilde{B}_{m} &= \sqrt{\frac{2G}{e^{2G\tau}-1}}\int_{0}^{\tau}dt\, b^{\rm in}_{m}(t)e^{(G-i\delta)t} ,\label{e8}
\end{align}
and use the input-output relation for the optical modes $a_{j}^{\rm out}=a_{j}^{\rm in}+\sqrt{2\kappa}a_{j}$ to show from equations~(\ref{e3})-(\ref{e7}) that {\it the output cavity modes behave collectively} and satisfy the following relations
\begin{align}
W_{\rm out} &= -\frac{G\!+\!\gamma}{G}\!\left[W_{\rm in}e^{G\tau} +i\sqrt{e^{2G\tau}-1}\, A^{{\rm in}\dag}_{m}\right] +\frac{\gamma}{G}\tilde{W}_{\rm in} ,\nonumber\\
U_{\rm out} &= -\tilde{U}_{\rm in} ,\label{e9}
\end{align}
where
\begin{align}
W_{\rm out} &= (\!\sqrt{G_{1}}A_{1}^{\rm out}\!+\!\sqrt{G_{2}}A_{2}^{\rm out}\!+\!i\sqrt{\gamma}\tilde{B}_{m}^{\dag})/\sqrt{G} ,\nonumber\\
\tilde{W}_{\rm in} &= (\!\sqrt{G_{1}}\tilde{A}_{1}^{\rm in}\!+\!\sqrt{G_{2}}\tilde{A}_{2}^{\rm in}\!-\!i\sqrt{\gamma}\tilde{B}_{m}^{\dag})/\sqrt{G} ,\nonumber\\
U_{\rm out} &= (\!\sqrt{G_{2}}A_{1}^{\rm out}\!-\!\sqrt{G_{1}}A_{2}^{\rm out}\!+\!i\sqrt{\gamma}\tilde{B}_{m}^{\dag})/\sqrt{G} ,\nonumber\\
\tilde{U}_{\rm in} &= (\!\sqrt{G_{2}}\tilde{A}_{1}^{\rm in}\!-\!\sqrt{G_{1}}\tilde{A}_{2}^{\rm in}\!-\!i\sqrt{\gamma}\tilde{B}_{m}^{\dag})\!/\sqrt{G} .\label{e10}
\end{align}
Clearly, the dynamics of the cavity modes involves only the collective ``symmetric'' mode $W$. The collective ``antisymmetric'' mode $U$ decouples from the remaining modes and does not evolve in time, i.e. $U$ is a constant of motion~\cite{wang12,dong12}. Thus even though the system is composed of three modes, the dynamics is fully determined in terms of only two modes, the mirror mode $A_{m}$ and the collective mode~$W$. In other words, the cavity modes act collectively on the mirror mode.

\textbf{Coherence between the cavity modes.} We first demonstrate that there is generally mutual coherence between the output cavity modes, crucial for the collective steering of the mirror. We assume that the cavity modes are in the ordinary vacuum state, the mirror mode is initially in a thermal state with the mean number of phonons $n_{0}$, and the oscillations of the mirror are subjected to the damping~$\gamma$ and to the Brownian noise~$n$. The mutual coherence between the output cavity modes is measured by the cross correlation $\langle A_{1}^{{\rm out} \dag}A_{2}^{\rm out}\rangle$ and with the help of equations~(\ref{e9}) and (\ref{e10}) we find
\begin{align}
|\langle A_{1}^{{\rm out} \dag}A_{2}^{\rm out}\rangle| &=\sqrt{\frac{G_{1}G_{2}}{G^{2}}}\Big[(n_{0}+1)\left(e^{2r}-1\right)+2(n+1) \frac{\gamma}{G} \frac{e^{2r} (\sinh2r - 2r)}{e^{2r}-1}\Big], \label{e11}
\end{align}
where $r = G\tau$ represents an effective ``squeezing parameter''. The cross correlation function is nonzero for $r>0$ indicating that the cavity modes, when driven by laser pulses tuned to the blue sideband of their average frequency, are generally mutually coherent. Notice a constructive role of the thermal noises present at the mirror, as well as the spontaneous decay $\gamma$ on the creation of the coherence.
\begin{figure}[h]
\centering{}\includegraphics[width=7.5cm]{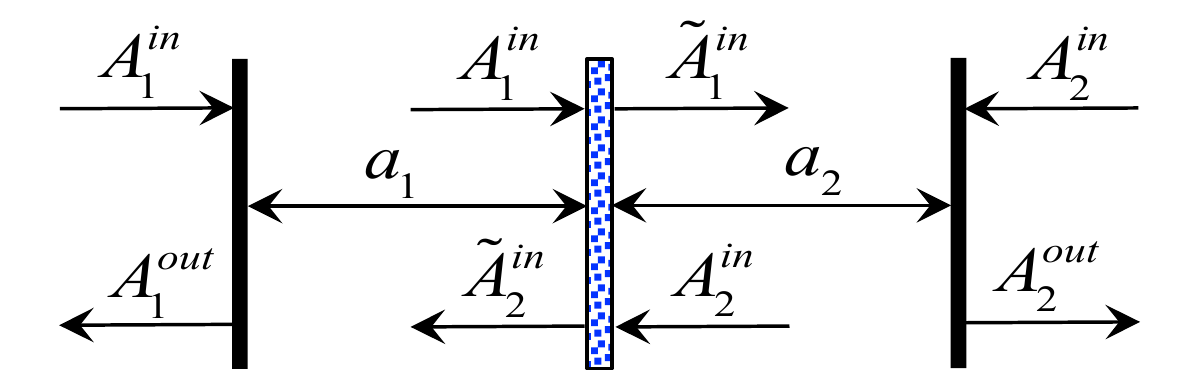}
\caption{\textbf{Schematic diagram of the input and output modes of the system.} The two cavity modes $a_1$ and $a_2$ exert a strong dynamical influence on one another via phase locking by the oscillating mirror which effectively acts as a partly reflecting and partly transmitting mirror.}
\label{fig2}
\end{figure}

The appearance of mutual coherence can be readily understood in terms of the phase relationship between the fields reflected and transmitted by the oscillating mirror. As illustrated in Fig.~\ref{fig2}, the system is equivalent to two single-mode two-sided cavities with the common oscillating mirror. We see that the mirror not only reflects but also transmits a part  $\tilde{A}_{j}^{\rm in}$ of the incident amplitude to the other mode. The phase reversal associated with reflection thus creates a phase difference between the reflected and transmitted amplitudes, hence fulfills the condition required for mutual coherence.

\textbf{Tripartite steering of the mirror mode.} We now turn to examine conditions for tripartite and collective steering of the mirror mode by the cavity modes. Quantum steering of mode $i$ by mode $j$ is normally identified by inferred quadrature variances defined as  $\Delta_{{\rm inf},j}^{2}X_{i}^{\rm out}=\Delta^{2}(X_{i}^{\rm out}+u_{j}O_{j}^{\rm out})$ and $\Delta_{{\rm inf},j}^{2}P_{i}^{\rm out}=\Delta^{2}(P_{i}^{\rm out}+u_{j}O_{j}^{\rm out})$, where $O_{j}^{\rm out}$ is an arbitrary quadrature for system $j$ and $u_{j}$ is a gain factor, both selected such that they {\it minimize} the inference (conditional standard deviation) product, $E_{i|j}=\Delta_{{\rm inf},j}X_{i}^{\rm out}\Delta_{{\rm inf},j}P_{i}^{\rm out}$. The observation of $E_{i|j}<1/2$ (with Heisenberg uncertainty relation satisfying $\Delta X_i\Delta P_i \geq \hbar/2, \hbar=1$) is an criterion that  $i$ can be steered by $j$, implying an EPR paradox~\cite{mr89,mr09}. Recent work of Wiseman, Jones, and Doherty~\cite{wj07} revealed that the EPR paradox is a realization of quantum steering. The correlation of $X_i, P_i$ with $X_j, P_j$ is strongest when $E_{i|j}=0$. This presents a perfect EPR state where measuring the position and momentum on system $j$ would provide a prediction, with $100\%$ accuracy, of the position and momentum of system $i$~\cite{mr89,mr09}. The quadratures are directly measurable in schemes involving homodyne or heterodyne detection.

Tripartite steering of the mirror by two cavity modes is determined by $E_{m|W} = \Delta_{{\rm inf},W}X_{m}^{\rm out}\Delta_{{\rm inf},W}P_{m}^{\rm out}= \Delta^{2}X_{m}^{\rm out} -\langle X^{\rm out}_{m},P^{\rm out}_{W}\rangle^{2}/\Delta^{2}P_{W}^{\rm out}$, where we define $\langle X^{\rm out}_{m},P^{\rm out}_{W}\rangle=\langle X^{\rm out}_{m}P^{\rm out}_{W} \rangle - \langle X^{\rm out}_{m}\rangle \langle P^{\rm out}_{W}\rangle$~\cite{mr89,mr09}, $X_{m}^{\rm out}=(A_{m}^{\rm out}+A_{m}^{{\rm out} \dag})/\sqrt{2}$ and $P_{W}^{\rm out}=(W_{\rm out}-W_{\rm out}^{\dag})/\sqrt{2}i$.  A value of $E_{m|W}$ smaller than $1/2$ indicates steering of the mirror mode $m$ by the collective mode $W$ of two optical fields, and the maximum EPR correlation presented by $E_{m|W}=0$ demonstrates a perfect EPR state, as explained above.

The tripartite steering parameter $E_{m|W}$ can be calculated analytically and in the limit of $\gamma\ll G$ is given by a simple analytical expression
\begin{align}
E_{m|W} = \left(n_{0}+\frac{1}{2}\right)\!\!\left[1-\frac{2\left(n_{0}+1\right)\!\left(e^{2r}-1\right)}{2\!\left(n_{0}\!+\!1\right)\!\left(e^{2r}\!-\!1\right)\!+\!1}\right].\label{e13}
\end{align}
This shows that $E_{m|W}$ is affected only by the initial thermal noise $n_{0}$ present at the mirror and $E_{m|W}$ is always smaller than $1/2$ when $n_{0}=0$. Even if $n_{0}\neq 0$, $E_{m|W}<1/2$ if $r>r_{\rm th}$, where $r_{\rm th}=\ln[(2n_0+1)/(n_0+1)]/2$ is a temperature-dependent minimal squeezing parameter which approaches to the limiting value of $\ln2/2$ as $n_0 \to \infty$. We see from equation~(\ref{e13}) that even if $n_{0}\neq 0$, one can always achieve a perfect EPR state, $E_{m|W}\rightarrow 0$, if $r\rightarrow\infty$, In other words, the mirror is always steered by the collective action of the cavity modes. We emphasize that the steering formula (\ref{e13}) is very general, it is independent of the frequency difference $\Delta$, the ratio~$G_{1}/G_{2}$, and the Brownian noise~$n$.

\textbf{Collective steering of the mirror mode.} To examine the occurrence of collective steering we must evaluate the bipartite steering parameter $E_{m|j}= \Delta_{{\rm inf},j}X_{m}^{\rm out}\Delta_{{\rm inf},j}P_{m}^{\rm out}=\Delta^{2}X_{m}^{\rm out} -\langle X^{\rm out}_{m},P^{\rm out}_{j}\rangle^{2}/\Delta^{2}P_{j}^{\rm out} $, and search for the condition $E_{m|j}\geq 1/2$. The quadrature of the $j$th cavity mode is defined as $P_{j}^{\rm out}=(A_{j}^{\rm out}-A_{j}^{\rm out\dag})/\sqrt{2}i$. With the help of equations~(\ref{e9}) and (\ref{e10}) and for $\gamma\ll G$ we may readily show that
\begin{align}
E_{m|1} = \frac{1}{2} + \frac{\left(G_{2}-G_{1}\right)\!\left(n_{0}+1\right)\!\left(e^{2r}-1\right)\!+\!n_{0}G}{2G_{1}\left(n_{0}\!+\!1\right)\!\left(e^{2r}\!-\!1\right)\!+\!G} .\label{e14}
\end{align}
The expression for $E_{m|2}$ can be obtained by simply interchanging $G_{1}\leftrightarrow G_{2}$. It is easily seen from equation~(\ref{e14}) that the bipartite steering of the mirror by either cavity modes is impossible when $G_{1}=G_{2}$. There is a simple physical interpretation of this feature by referring to the monogamy condition that two modes cannot simultaneously steer an another mode~\cite{monogamy}. When $G_{1}=G_{2}$ the two cavity modes are unresolved at the mirror. Then, it is impossible to tell which of the modes steers the mirror mode. Since $E_{m|1}=E_{m|2}\geq 1/2$ over the entire range of $r$, there are no restrictions on steering of the mirror collectively by the cavity modes. We therefore conclude that the system can be considered as an example of a practical system where the secure multimode protocol can be realized. The mirror mode (Alice) can be collectively steered by the cavity modes (Bob and Charlie).

\begin{figure}[h]
\center{}\includegraphics[width=7cm]{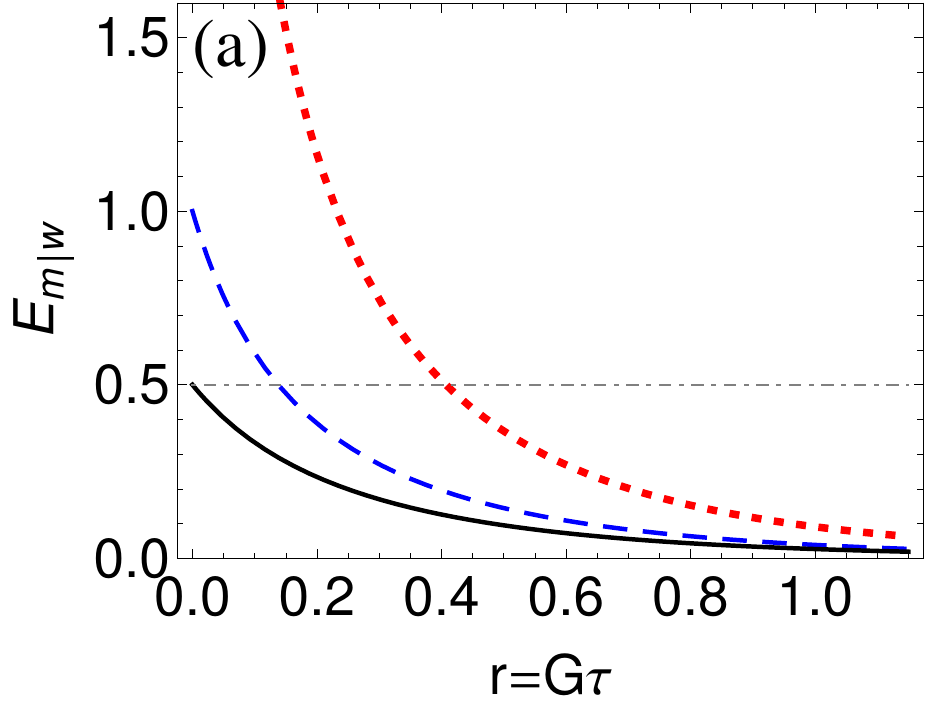}
\includegraphics[width=7cm]{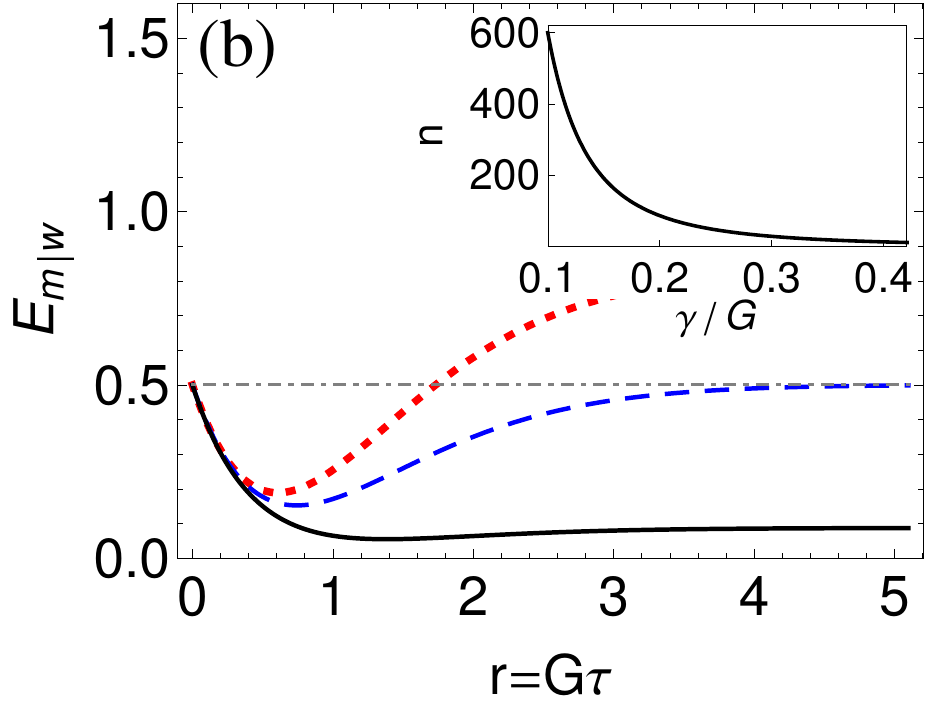}
\caption{\textbf{The robustness of the collective steering against thermal noises.} Collective steering parameter $E_{m|W}$ as a function of $r$ for different values of $n_0$ and $n$. Parameters used here are $\gamma/G = 0.1$, (a) $n_{0}=n=0$ (solid), $n_{0}=n=0.5$ (dashed), $n_{0}=n=5$ (dotted); (b) $n_{0}=0$ and $n=100$ (solid), $n=600$ (dashed), $n=1000$ (dotted). The inset of (b) shows the threshold of $n$ for each $\gamma/G$, below that collective steering can occur over the entire range of $r$. }
\label{fig3}
\end{figure}

\textbf{Effect of losses and experimental feasibility.} We consider the effect of losses, the thermal noises $n_{0}$ and $n$, as well as the damping $\gamma$ on the tripartite steering. From Figs.~\ref{fig3}(a) and \ref{fig3}(b), we see that in general the thermal noise has a destructive effect on the steering. However, a near perfect steering can still be observed for a relatively large $n\sim100$. For even larger $n$ below a threshold, collective steering can still occur over the entire range of $r$. As shown in the inset of Fig.~\ref{fig3}(b), this threshold can be as high as $n\sim 600$ for $\gamma/G=0.1$, and decreases with increasing damping. As $n$ further increases, collective steering can only take place in a restricted range of $r$.

\begin{figure}[h]
\center{}\includegraphics[width=9cm]{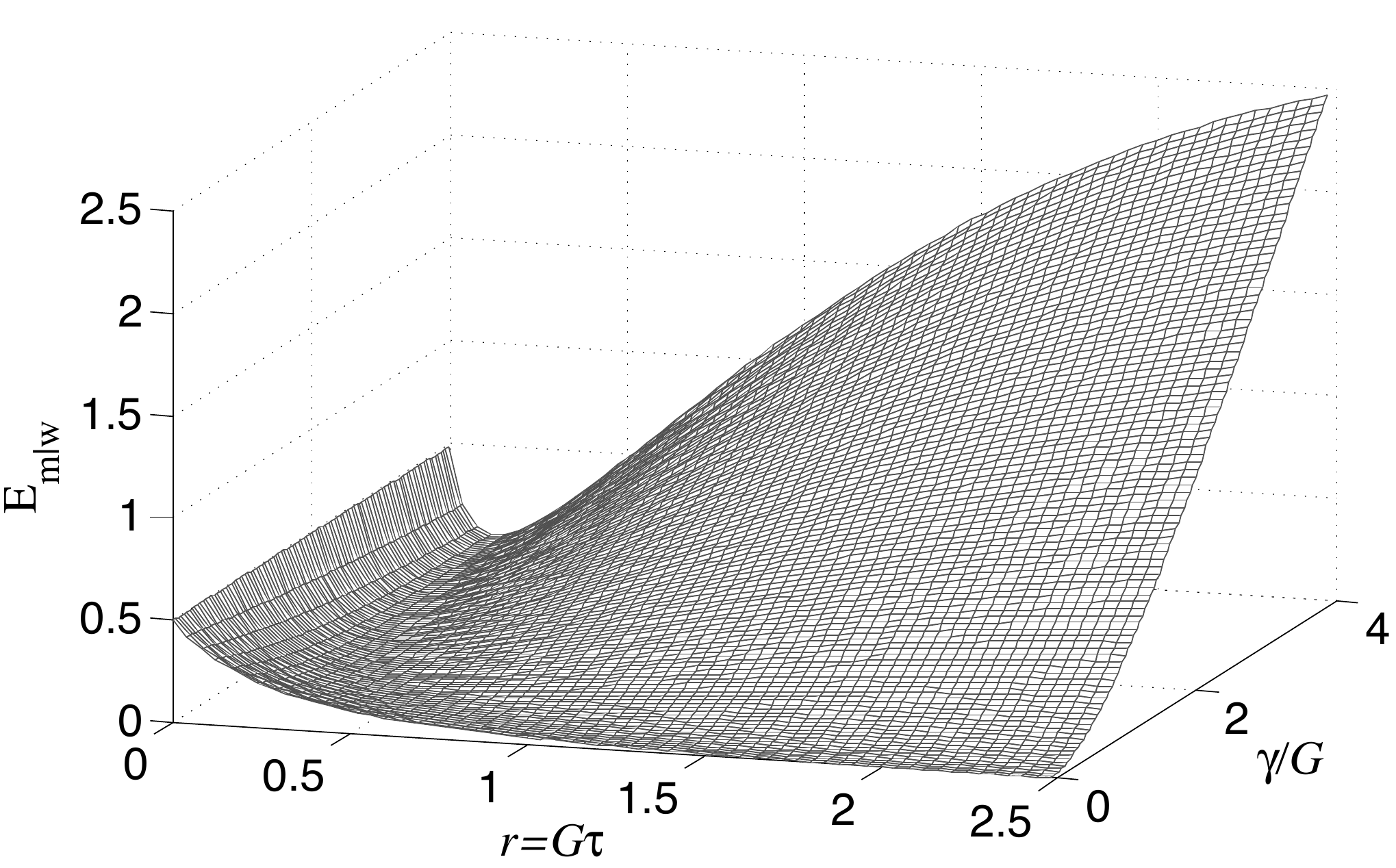}
\caption{\textbf{The effect of mechanical damping.} Collective steering parameter $E_{m|W}$ as a function of $r$ and $\gamma/G$ for $n_{0}=n=0$.}
\label{fig4}
\end{figure}

The robustness of the collective steering against the damping $\gamma$ is illustrated in Fig.~\ref{fig4}. From this result, we notice that collective steering is present in a large parameter regime except for the up right corner where both $r$ and $\gamma/G$ are large. It is interesting and somewhat surprising that in the presence of the mirror damping, the collective steering is preserved at small $r$ rather than at large~$r$. This can be understood by recalling that a strong squeezing results in a large sensitivity of the variances to the external noise.

We emphasize that the requirements for thermal noise and damping rate to achieve collective steering in the present system are attainable in existing experiments. In particular, in a recent experiment on nanoscale optomechanics~\cite{chan11}, the mechanical oscillator was cooled down to a temperature corresponding to $n_{0} =  n = 0.85$, with frequency $\omega_{m}/2\pi = 3.68$ GHz and damping rate of $\gamma/2\pi=35$ kHz. A typical driving pulse laser of $\lambda_{L}=1537$ nm corresponds to an optical line width of $\kappa/2\pi = 500$ MHz. The effective opto-mechanical coupling constant $g/2\pi=40.7$ MHz. These give $G_1=G_2 =g^2\kappa/(\kappa^2+\Delta^2)=2\pi\times1.66$ MHz (assumed $\Delta = \kappa$), and $\gamma/G\approx1.05\times10^{-2}$. The effect of the thermal noise and the damping on the collective steering is therefore expected to be small in general. In an another experiment in which the EPR entanglement between mechanical motion and a microwave field has been demonstrated~\cite{lehnert13}, the noises and the damping rate were estimated as $n_0 = 0.5$, $n = 37.8$, and $\gamma/G \approx 1.75 \times 10^{-3}$, respectively. These values are also well within the safety zone for a successful collective steering.  Therefore, the perfect collective EPR steering we have been discussing should be observable with current experiments.

\section*{Discussion}
We have proposed an efficient scheme for the preparation of a perfect EPR state of the macroscopic mirror and the collective mode of two optical fields in an optomechanical cavity. By tuning the laser pulses to the blue sideband of the average frequency of cavity modes, we have demonstrated that the oscillating mirror has the effect of inducing a coherence between the modes which is crucial for collective steering of the mirror. We also demonstrate the robustness of the collective steering against thermal noises and mechanical damping, and conclude that our scheme can be readily implemented with the existing experimental techniques. The collective steering is a newly encountered feature which opens promising perspective for the realization of secure multi-mode quantum cryptography and perfect one-sided device-independent quantum secret sharing.

\section*{Methods}
\textbf{The derivation of Langevin equation (1).} The full Hamiltonian for the system in the interaction picture, including the laser driving term and the nonlinear radiation pressure interaction, is given by
\begin{align}
H_{I} &=\hbar\triangle_{01}a_{1}^{\dagger}a_{1}+\hbar\triangle_{02}a_{2}^{\dagger}a_{2}+\hbar\omega_{m}b_{m}^{\dagger}b_{m}+\hbar g_{01}a_{1}^{\dagger}a_{1}(b_{m}+b_{m}^{\dagger})+\hbar g_{02}a_{2}^{\dagger}a_{2}(b_m+b_m^{\dagger})+iE_{1}(a_{1}^{\dagger}-a_{1})+iE_{2}(a_{2}^{\dagger}-a_{2}),
\end{align}
where $\triangle_{0j}=\omega_{j}-\omega_{L}\ (j=1,2)$ is the detuning (for the case of a cavity with fixed length) of the laser frequency from the frequency of the $j$th cavity mode, $g_{0j}$ is the single-photon coupling strength of the $j$th cavity mode to the mirror.

The quantum Langevin equations for the annihilation operators of the cavity modes $a_{j}$, the position $x_{m}$ and momentum $p_{m}$ operators of the mirror, are given by
\begin{align}
\dot{a}_{j} &= -\left(\kappa_{j} +i\Delta_{0j}\right)a_{j} -i\!\sqrt{2}g_{0j}a_{j}x_{m} +E_{j}-\sqrt{2\kappa_{j}}a_{j}^{\rm in} ,\nonumber\\
\dot{p}_{m} &= -\gamma p_{m}-\omega_{m}x_{m}-\!\sqrt{2}(g_{01}a_{1}^{\dag}a_{1}+g_{02}a_{2}^{\dag}a_{2})-\!\sqrt{2\gamma}\xi ,\nonumber\\
\dot{x}_{m} &= \omega_{m}p_{m} ,\label{e1}
\end{align}
where $x_{m}=(b_{m}+b_{m}^{\dagger})/\sqrt{2},\ p_{m}=(b_{m}-b_{m}^{\dagger})/\sqrt{2}i$ are the position and momentum operators of the mechanical oscillator, $\kappa_{j}$ and $\gamma$ are, respectively, the damping rate of the cavity mode $j$ and the mirror. $a_{j}^{\rm in}$ and $\xi$ are Langevin noise terms which are taken to be statistically independent with nonzero correlated functions $\langle a_{j}^{\rm in}(t)a_{j}^{\rm in\dag}(t^{\prime})\rangle =\delta(t-t^{\prime})$ (in the optical vacuum state), and $\langle\xi(t)\xi(t^{\prime}) +\xi(t^{\prime})\xi(t)\rangle = (2n+1)\delta(t-t^{\prime})$, where~$n$ is the mean number of the thermal phonons of the mechanics.

Next, we write each operator in equation~(\ref{e1}) as a $c$-number plus a fluctuation operator, $a_{j} = \alpha_{j}+\delta a_{j}$, $p_{m}=p_{s}+\delta p_{m},$ $x_{m}=x_{s}+\delta x_{m}$, where $\alpha_{j}=E_j/(\kappa_{j}+i\Delta_{j}),\ p_{s}=0,\ x_{s}=-\sqrt{2}(g_{1}|\alpha_{1}|^{2}+g_{2}|\alpha_{2}|^{2})/\omega_m$ are the mean amplitudes of the cavity fields, the momentum and position of the mirror. $\Delta_{j}=\Delta_{0j}+\sqrt{2}g_{0j}x_{s}$ is the effective detuning including the radiation pressure effects, $g_{j}=g_{0j}|\alpha_{j}|$ is the effective coupling strength of the $j$th cavity mode to the mirror. We then arrive to the following linearized Langevin equations for the fluctuation operators
\begin{align}
\delta \dot {a_{j}} &=  -(\kappa_{j}+i\Delta_{j})\delta a_{j}-i\sqrt{2}g_{j}\delta x_{m}-\sqrt{2\kappa_{j}}a_{j}^{in},\nonumber \\
\delta \dot{p_{m}} &=  -\gamma\delta p_{m}-\omega_{m}\delta x_{m}-\sqrt{2}g_{1}(\delta a_{1}^{\dagger}+\delta a_{1})-\sqrt{2}g_{2}(\delta a_{2}^{\dagger}+\delta a_{2})-\!\sqrt{2\gamma}\xi,\nonumber \\
\delta  \dot{x_{m}}  &= \omega_{m}\delta p_{m}.
\end{align}

Denoting the average frequency of the cavity modes by $\omega_{0}=(\omega_{1}+\omega_{2})/2$ and the frequency difference by $\Delta =(\omega_{1}-\omega_{2})/2$, we focus on the case of~$\omega_{L}=\omega_{0}+\omega_{m}$, i.e. the laser frequency $\omega_{L}$ on resonance with the blue sideband of the average frequency $\omega_{0}$ of the cavity modes. Consider a frame rotating with $\omega_m$ by substituting $\delta{a}_{j}\rightarrow \delta a_{j} e^{-i\omega_{m}t}$, $\delta{a}_{j}^{in}\rightarrow \delta a_{j}^{in} e^{-i\omega_{m}t}$, $\delta{b}_{m}\rightarrow \delta b_{m}e^{i\omega_{m}t}$, in the limit $g_j\ll \kappa_j \ll \omega_m$, we then obtain the Langevin equation (1) for the fluctuation operators within rotating-wave approximation.

\textbf{The solutions for the variances of the output modes.} The expressions for the bipartite and tripartite steering parameters are in terms of the variances and correlation functions. Here, we give out the quadrature components of the output fields in terms of the quadrature components of the input fields. For light initially in vacuum $\Delta^{2}X_{j}^{in}=\Delta^{2}P_{j}^{in}=1/2,\ \Delta^{2}\tilde{X}_{j}^{in}=\Delta^{2}\tilde{P}_{j}^{in}=1/2$ and the mirror in a thermal state $\Delta^{2}X_{m}^{in}=\Delta^{2}P_{m}^{in}=n_{0}+1/2$, we arrive to the following expressions for the variances of the output fields:
\begin{align}
\Delta^{2}X_{m}^{out} & = \Delta^{2}P_{m}^{out}\\
&= -\frac{1}{2}+(n_{0}+1)e^{2r}+\frac{\gamma}{G}(n+1)(e^{2r}-1),\nonumber \\
\Delta^{2}X_{j}^{out} & = \Delta^{2}P_{j}^{out}\\
&= \frac{1}{2}+\frac{G_{j}}{G}(n_{0}+1)(e^{2r}-1)+\frac{G_{j}\gamma}{G^{2}}(n+1)(e^{2r}+1-\frac{4re^{2r}}{e^{2r}-1}),\nonumber \\
\Delta^{2}X_{W}^{out}&=\Delta^{2}P_{W}^{out}\\
&=(1+\frac{\gamma}{G})^{2}\{[n_{0}+1+\frac{\gamma}{G}(n+1)]e^{2r}-(n_{0}+\frac{1}{2})\} +\frac{\gamma}{G}[\frac{1}{2}+\frac{\gamma}{G}(n+1)][\frac{\gamma}{G}-(1+\frac{\gamma}{G})\frac{4r e^{2r}}{e^{2r}-1}],\nonumber \\
\langle X_{m}^{out},P_{j}^{out}\rangle  & =\langle P_{m}^{out},X_{j}^{out}\rangle \\
&= -\sqrt{\frac{G_{j}}{G}}\sqrt{e^{2r}-1}e^{r}[n_{0}+1+\frac{\gamma}{G}(n+1)(1-\frac{2r}{e^{2r}-1})],\nonumber \\
\langle X_{m}^{out},P_{W}^{out}\rangle &=\langle P_{m}^{out},X_{W}^{out}\rangle \nonumber \\
&= -(1+\frac{\gamma}{G})e^{r}\sqrt{e^{2r}-1}[n_{0}+1+\frac{\gamma}{G}(n+1)]+\frac{\gamma}{G}\frac{2r e^{r}}{\sqrt{e^{2r}-1}}[\frac{1}{2}+\frac{\gamma}{G}(n+1)].
\label{BB_xp}
\end{align}

\section*{Acknowledgements}
Q.Y.H. thanks support from the National Natural Science Foundation of China under grants (Nos. 11274025 and 61475006). Q.H.G thanks support from the National Natural Science Foundation of China under grants (No. 11121091).

\section*{Author contributions statement}
M.W., Q.Y.H and Z.F conceived the idea, Z.F and M.W. performed the calculations, Q.Y.H and Z.F analyzed the results and wrote the main manuscript text, Q.H.G participated in the discussions. All authors reviewed the manuscript.

\section*{Additional information}
Competing financial interests: The authors declare no competing financial interests.

\end{document}